\documentclass[final]{raa06}           
\usepackage{graphicx,times}             
\usepackage{natbib}
\usepackage{amssymb,amsmath}
\bibpunct{(}{)}{;}{a}{}{,}
\usepackage[colorlinks=true, citecolor=blue]{hyperref}%

\usepackage{graphicx,kantlipsum,setspace}
\usepackage{caption}
\usepackage{graphics,epsf}
\usepackage{amsmath}                
\usepackage{amsfonts}               
\usepackage{amssymb}                
\usepackage{epsfig}                 
\usepackage{appendix}
\usepackage{graphicx}
\usepackage{float}
\usepackage{color}
\usepackage{multirow}
\usepackage{colortbl}
\usepackage[para,online,flushleft]{threeparttable}
\usepackage{xcolor}

\hypersetup{citecolor=blue, 
            linkcolor=red, 
            menucolor=blue, 
            urlcolor=blue}  

\newcommand{\km}{{~\rm km}}

\begin{document}

   \title{Hints of point-symmetric structures in SN 1987A: the jittering jets explosion mechanism
}

   \volnopage{Vol.0 (20xx) No.0, 000--000}      
   \setcounter{page}{1}          

   \author{Noam Soker
    }

   \institute{Department of Physics, Technion, Haifa, 3200003, Israel;   {\it   soker@physics.technion.ac.il}\\
\vs\no
   {\small Received~~20xx month day; accepted~~20xx~~month day}}

\abstract{I identify a point-symmetric structure composed of three pairs of clumps in the recently released JWST image of the ejecta of SN 1987A and argue that these pairs of clumps support the jittering jets explosion mechanism (JJEM) for SN 1987A. I compare this point-symmetric structure with the multipolar-lobe morphology of a post-asymptotic giant branch nebula. The three pairs of clumps in the post-AGB nebula are formed at the tip of jet-inflated lobes.
I use this similarity to strengthen earlier claims that SN 1987A was exploded by jets in the frame of the JJEM.  
\keywords{stars: massive -- stars: neutron -- supernovae: general -- stars: jets -- ISM: supernova remnants --  supernovae: individual (SN 1987A)}}

 \authorrunning{N. Soker}            
\titlerunning{The point-symmetric structure of SN 1987A}  
   
      \maketitle

\section{Introduction} 
\label{sec:intro}

The jittering jets explosion mechanism (JJEM) of core collapse supernovae (CCSNe) is based on pairs of jets with stochastic varying directions that the newly born neutron star (NS) launches in the first second to several seconds of the explosion process of massive stars (e.g., \citealt{Soker2010, PapishSoker2011, GilkisSoker2015,  Soker2020RAA, ShishkinSoker2021,  Soker2022SNR0540, Soker2023gap}). The last jets to be launched account for `ears' in supernova remnants (SNRs; e.g., \citealt{,Bearetal2017, GrichenerSoker2017, YuFang2018}). 

The source of the stochastically varying angular momentum of the mass that the NS accretes results from pre-collapse stochastic core convection motion (e.g., \citealt{Soker2010, PapishSoker2014Planar, GilkisSoker2015, Soker2019SASI, ShishkinSoker2021, ShishkinSoker2022, Soker2022SNR0540, Soker2022Boosting}; or envelope convection motion if a black hole is formed, e.g., \citealt{Quataertetal2019, AntoniQuataert2022, AntoniQuataert2023}) that are amplified by instabilities behind the stalled shock at $\simeq 100 \km$ from the NS. In cases where the pre-collapse core rotation is non-negligible, the stochastic angular momentum variations are around the angular momentum axis of the pre-collapse core (e.g., \citealt{Soker2023gap, Soker2023Classes}).

A strong and very clear prediction of the JJEM is that in some (but not all) cases the jets imprint point-symmetrical morphological components in the ejecta. Point-symmetry refers to two similar  structural components on opposite sides of the center of the explosion. Because CCSNe are highly non-spherical the two opposite structural components of a pair might have a different small-scale structure, a different brightness, and/or be at a different distance from the center. Examples of point-symmetry morphologies in CCSN remnants that support the JJEM are the velocity maps of SNR 0540-69.3 (\citealt{Soker2022SNR0540}; observations by \citealt{Larssonetal2021}) and the images of SNR Vela (\citealt{Soker2023Classes}; observations by, e.g., \citealt{Aschenbachetal1995, Sapienzaetal2021}.).   

There is a dispute on whether the non-spherical structural features of the ejecta of SN 1987A, which include clumps and filaments (e.g., \citealt{Franssonetal2015, Franssonetal2016, Larssonetal2016, Abellanetal2017, Matsuuraetal2017, Larssonetal2023}), are due only to instabilities during the explosion that was powered by the delayed neutrino mechanism (e.g, \citealt{Kjaeretal2010, Jerkstrandetal2020}), or whether jets in the frame of the JJEM also played a role in shaping the ejecta (e..g, \citealt{Soker2017TwoPromissing, BearSoker2018SN1987A}), in addition to instabilities that exist also in the JJEM. 
\cite{Onoetal2020} and \cite{Orlandoetal2020} conclude, based on their 3D hydrodynamical simulations, that jet-driven explosion best reproduce the explosion morphology and element distribution. They, however, took the jets' axis in the plane of the inner circumstellar ring, while \cite{BearSoker2018SN1987A} took the main jets' axis to be at an angle to the plane of the inner ring.

\begin{figure*}[th]
	\centering
\includegraphics[trim=2.8cm 12.1cm 3.0cm 2.4cm ,clip, scale=0.52]{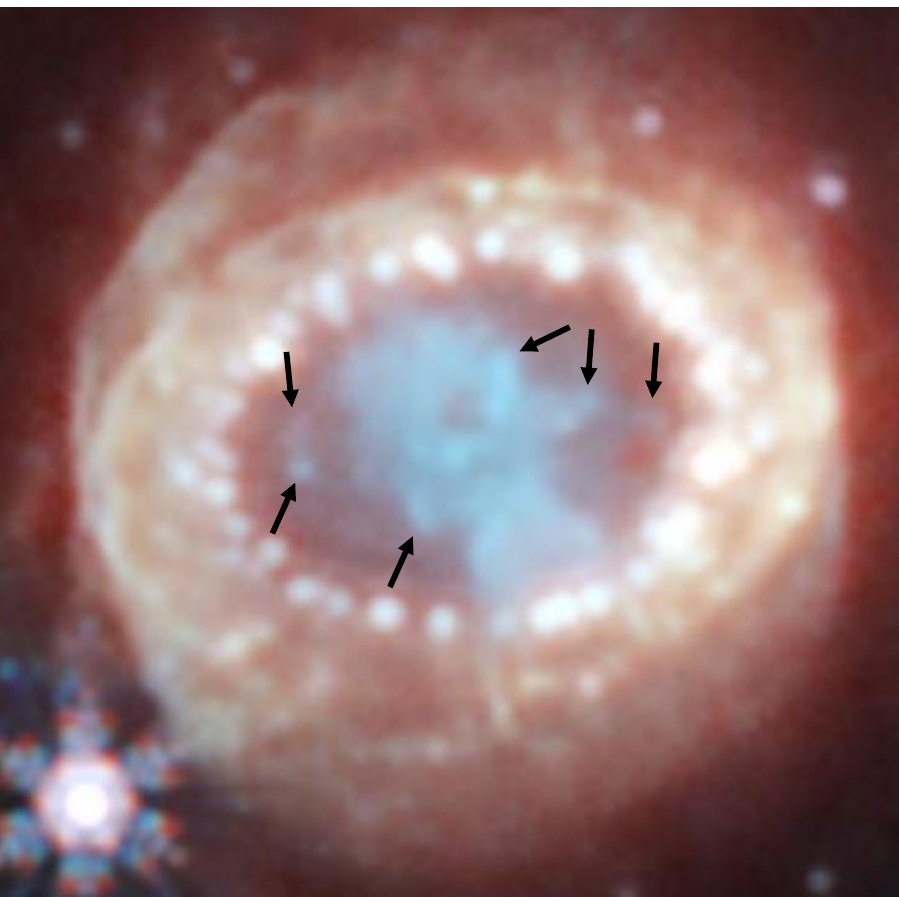} 
\includegraphics[trim=2.8cm 12.1cm 3.0cm 2.4cm ,clip, scale=0.52]{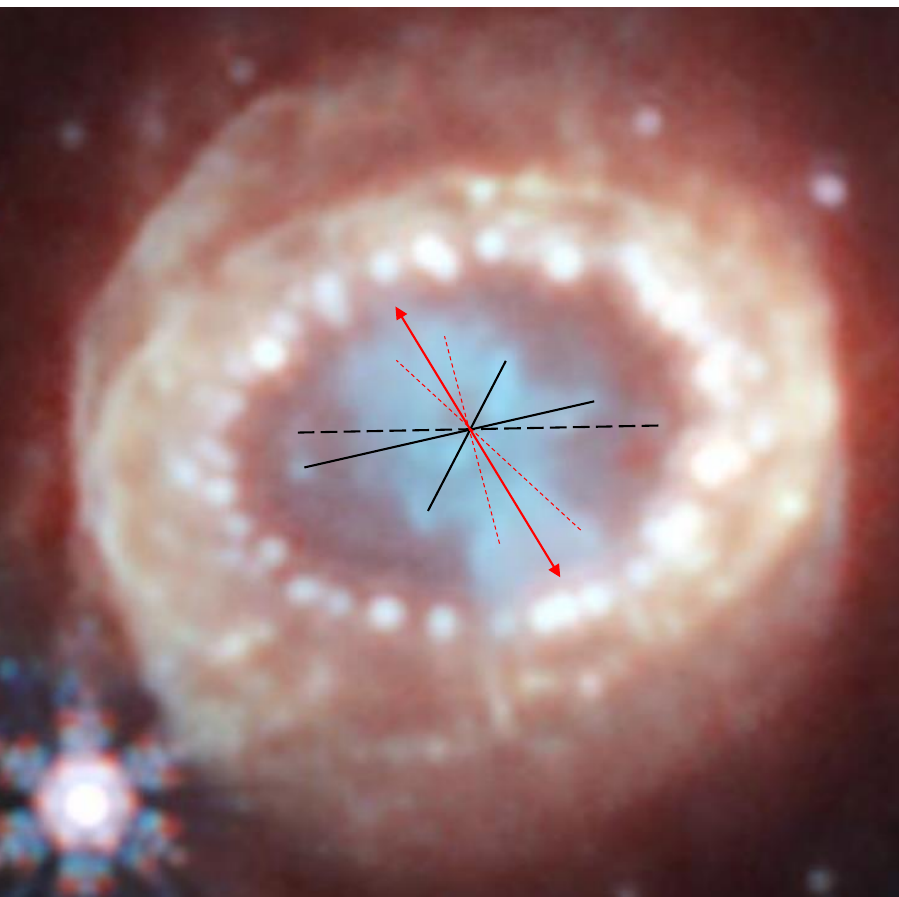} \\
\includegraphics[trim=1.2cm 18.2cm 1.9cm 3.9cm ,clip, scale=1.1]{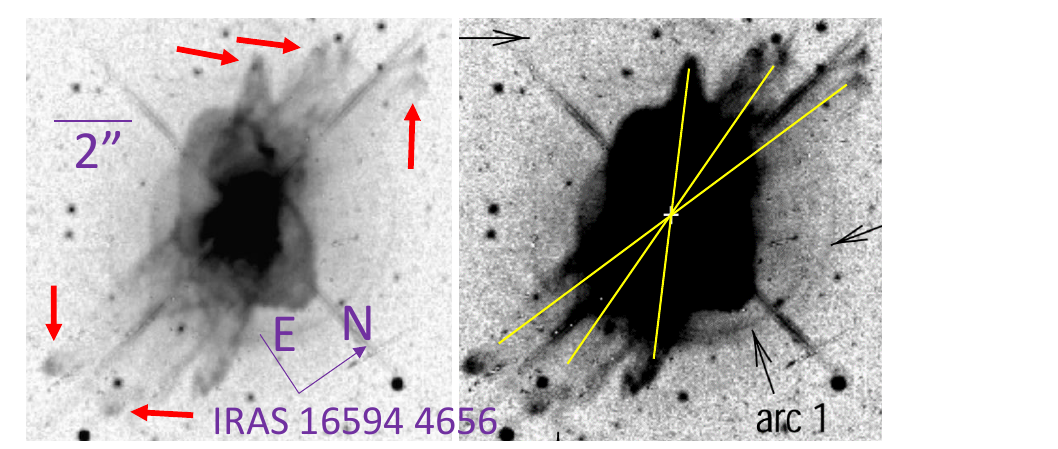} 
\caption{Upper panels: A JWST image of SN 1987A (Credits: NASA, ESA, CSA, M. Matsuura (Cardiff University), R. Arendt (NASA’s Goddard Spaceflight Center \& University of Maryland, Baltimore County), C. Fransson). 
Blue represents emission at 1.5 microns, cyan at 1.64 and 2.0 microns, yellow at 3.23 microns, orange at 4.05 microns (F405N), and red at 4.44 microns. 
Upper left: The arrows I added point at some regions on the outskirts of the blue-cyan structure that are brighter than their surroundings, i.e., clumps. 
Upper right: I added three lines that connect pairs of clumps as I marked in the upper left panel. The double-headed arrow marks the direction of the elongated axis of the blue-cyan zone. Red-dashed lines indicate possible jet-related structures. 
Lower left: An HST image of the post-AGB star IRAS 16594-4656 based on \cite{Hrivnaketal1999} and \cite{Hrivnaketal2001} where more details can be found. I added five red arrows to point at clumps. Each clump is at the tip of a lobe, suggesting it was shaped by a jet. Lower right: An HST image (F814W filter) from \cite{Hrivnaketal2001}. I added three yellow lines to mark pairs of clumps. The center of each line is at the center of the white-plus sign (where the lines cross each other), showing that two clumps of a pair can be at different distances from the center.   
}
\label{Fig:Fig1}
\end{figure*}

In this Letter I argue that the new JWST images of the SN 1987A remnant (SNR 1987A) hint at point-symmetry morphological features as expected to be shaped by jets in the frame of the JJEM. In Figure \ref{Fig:Fig1} I present the JWST image of SN 1987A as was released recently by Matsuura et al. (2023)\footnote{https://www.nasa.gov/feature/goddard/2023/webb-reveals-new-structures-within-iconic-supernova}. 
On that image I point with arrows at six clumps in the outer parts of the inner ejecta (upper left panel). 
By a clump I refer to a small region that is brighter than its surroundings, in any of the filters. On the image on the upper right panel I connect three pairs with black lines to mark the point-symmetry of the clumps. I suggest that these clumps were formed at the outer tip of the lobes that jets inflated during the late time of the explosion. Early jittering jets in the JJEM explode the core and might not leave morphological signatures (e.g., \citealt{Soker2023Classes}).
I also added two red-dashed lines to indicate possible jet-related structures and the red double-headed arrow for the long axis of the inner ejecta (the blue-cyan region). 

My claim that jets can form point-symmetric clumps as seen in SN 1987A is based in part on the morphologies of some planetary nebulae (PNe) and proto-PNe, namely, those with multipolar lobes (e.g., \citealt{Sahai2000}). Point-symmetry is a widespread characteristic of proto-PN and PN morphologies and are formed by the action of jets  
(e.g., \citealt{SahaiTrauger1998, Sahaietal2007, Sahaietal2011}). 
One example is the post-AGB star IRAS 16594-4656 \citep{Hrivnaketal1999}. 
I take an image based on \cite{Hrivnaketal1999} 
 and \cite{Hrivnaketal2001}
\footnote{For a newer image composed by Judy Schmidt, see: 
https://www.flickr.com/photos/geckzilla/11284828416} and present it in the lower left panel of Figure \ref{Fig:Fig1}. I mark some (but not all) clumps by red arrows. On the lower right panel I connect three pairs of clumps. Such proto-PNe and PNe are thought to be shaped by jets (e.g., \citealt{Morris1987, Soker1990AJ, SahaiTrauger1998, Boffinetal2012}). 
The structure of IRAS 16594-4656, and of many others proto-PNe (e.g., \citealt{Sahaietal2000}), show the following. (1) Jets can inflate lobes that form clumps at their tip. (2) The two opposite clumps of a pair might have different, structure, brightness, and distance from the center. (3) Other clumps exist in the system. 
 
In the JWST image of SNR 1987A I also find other clumps and that two clumps of a pair are not equal in their properties. 
The comparison of the point-symmetric structures of SNR 1987A with that of IRAS 16594-4656 suggests that jets also shaped the clumps of SNR 1987A that I marked on the upper panels. If holds, this strongly support the JJEM for SN 1987A.  

I note that surrounding magnetic field might shape only one pair of ears (e.g., \citealt{Wuetal2019}) and might play other roles in SNRs (e.g., \citealt{Xiaoetal2022}). However, magnetic fields cannot explain a point symmetric structure. 
Surrounding density inhomogenieties might also shape SNRs (e.g., \citealt{Luetal2021}), but not to form point-symmetric structure in the inner ejecta. 

I encourage deep imaging of SNR 1987A to further examine point-symmetry and related jet-shaped structural components. Any future analysis of the explosion of SN 1987A to consider JJEM. 





\label{lastpage}

\end{document}